\documentclass{emulateapj} 
\usepackage{latexsym}
\usepackage{mathrsfs}
\usepackage{times} 
\usepackage{graphics}
\usepackage[german,english,american]{babel} 

\begin{document}
\shorttitle{Heating the bubbly ICM}
\title{Heating the bubbly gas of galaxy clusters with weak shocks and
sound waves}

\shortauthors{Heinz \& Churazov}
\author{S.~Heinz$^{1,2}$, E.~Churazov$^{3,4}$\\$^{1}$
  Center for Space research, MIT, 77 Mass. Ave., Cambridge, MA
  02139\\$^{2}$ Chandra Fellow\\$^{3}$ Max-Planck-Institute for
  Astrophysics, Karl-Schwarzschild-Str.~1, 85741 Garching,
  Germany\\$^{4}$ Space Research Institute (IKI), Profsoyuznaya 84/32,
  Moscow 117810, Russia}

\begin{abstract}
Using hydrodynamic simulations and a technique to extract the
rotational component of the velocity field, we show how bubbles of
relativistic gas inflated by AGN jets in galaxy clusters act as a
catalyst, transforming the energy carried by sound and shock waves to
heat.  The energy is stored in a vortex field around the bubbles which
can subsequently be dissipated. The efficiency of this process is set
mainly by the fraction of the cluster volume filled by (sub-)kpc scale
filaments and bubbles of relativistic plasma.
\end{abstract}
\keywords{hydrodynamics --- instabilities --- shock waves --- methods:
  numerical --- galaxies: clusters: general --- ISM: bubbles}

\section{Introduction}
\label{sec:intro}
Models for the formation of galaxy clusters require an external source
of energy to avoid catastrophic cooling of gas at the centers of
massive clusters and to explain the observed lack of cool gas
\citep{peterson:03} in so-called cooling flow clusters
\citep{fabian:94}.  Relativistic jets from supermassive black holes
are the most likely source of this energy \citep{binney:95}.  While
jets may carry sufficient kinetic energy, it has been unclear how
efficiently that energy can be transferred to heat the cluster gas.

When jets interact with the gas in galaxy clusters (the intra-cluster
medium, ICM), they invariably inflate hot, tenuous bubbles of
relativistic gas, observed as synchrotron radio nebulae
\citep{scheuer:74}, a.k.a. radio lobes.  Evidence is mounting
\citep{ensslin:99,birzan:04,giovannini:04} that the ICM is littered
with such bubbles, each of which starts out growing supersonically,
driving a strong shock \citep{heinz:98} that efficiently heats the ICM
it encounters.  However, X-ray observations show that the expansion
rapidly decelerates and most bubbles expand subsonically, therefore
not producing strong shocks \citep{fabian:00,mcnamara:00}.
 
The bubbles themselves store a fraction of the initial jet energy,
some of which can be transferred to the gas and dissipated during the
bubble's slow, buoyant rise in the cluster atmosphere
\citep{churazov:01,begelman:01}.  Another significant fraction of the
jet energy is released in the form of sound and weak shock waves
\citep{fabian:03b}.  In homogeneous media, dissipation of wave energy
is determined by microscopic transport processes like viscosity or
conduction, which are well known for unmagnetized plasmas.
Realistically, though, the presence of even weak $B$-fields in the ICM
can alter the transport coefficients, making estimates of the
dissipation rate very uncertain.  High values for the microscopic
viscosity have been postulated as a way to dissipate the wave energy
\citep{ruszkowski:04,reynolds:05,fabian:05}.

In this letter we argue that the presence of filaments and bubbles of
relativistic gas provides an efficient way to extract the wave energy
and heat the ICM {\em even if} viscosity and thermal conduction are
strongly suppressed.  In \S\ref{sec:richtmyer} we review the process
responsible for this dissipation, \S\ref{sec:numerical} presents the
numerical methods we used, \S\ref{sec:results} discusses how our
results can be applied to cluster heating, and \S\ref{sec:conclusions}
summarizes.

\section{The Richtmyer-Meshkov Instability}
\label{sec:richtmyer}
The energy transfer to the ICM occurs through a process known as the
Richtmyer-Meshkov instability
\citep[RMI][]{richtmyer:60,meshkov:69,quirk:96,inogamov:99}.  A plane
shock or sound wave passing over a density discontinuity that is
inclined with respect to the wave front, as it occurs at the boundary
of the ICM with a relativistic bubble, induces a well-localized vortex
flow around the bubble \citep{ensslin:02}, as illustrated in Fig.~1.
A qualitatively similar process occurs when a shock passes over {\em
overdense} clouds \citep{klein:94,kornreich:00}.
\begin{figure}
\begin{center}
\resizebox{\columnwidth}{!}{\includegraphics{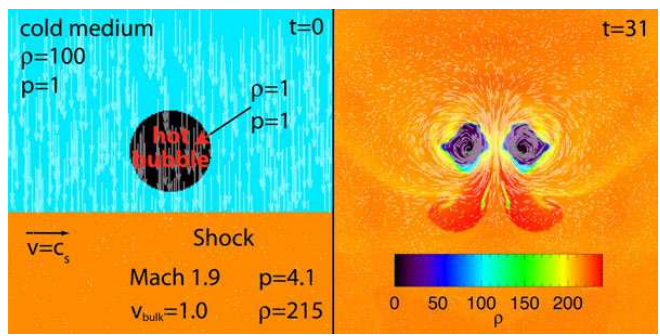}}
\end{center}
\caption{Map of fluid density (color scale) and velocity field
  (arrows) for Mach 1.9 shock running over a bubble with density
  contrast of 1:100. Shown are time frames 0 and 31 in units of the
  sound crossing time of the bubble.\label{fig:dens}}\vspace*{-12pt}
\end{figure}

Because the high sound speed keeps the material inside the bubble at
uniform (but time varying) pressure, the bubble acts like a hydraulic
piston, distributing pressure applied at one end over its entire
surface.  The arrival of the wave sets up a pressure gradient which
accelerates the bubble/wave interface inward, introducing significant
{\em inward} transverse motion and distorting the bubble.  {\em Ahead}
of the wave, the overpressured bubble expands into the ICM, inducing
significant {\em outward} transverse motion.  Once the shock front has
passed over the bubble, it leaves behind a vortex ring in what was
previously irrotational flow.  For a compression wave of finite width
$\lambda$, the passage of the trailing end of the wave will act in the
opposite sense, partly restoring the bubble's original shape and
canceling previously generated vorticity.  This reduces the amount of
energy deposited by the wave if $\lambda$ is not large compared to the
bubble radius $R$.

In the limit of low bubble density $\rho_{\rm bubble} \ll \rho_{\rm
ICM}$, and for $R\ll\lambda$, one can easily estimate the amount of
the energy funneled into this vortex: Material of density $\rho_{\rm
ICM}$ enters the bubble volume $V_{\rm bubble}$ with velocity $v_{\rm
ICM}$ (the bulk velocity of the compressed gas).  Thus, the energy
trapped in the toroidal vortex should be
\begin{equation}
E_{\rm rot}=g\frac{1}{2}V_{\rm bubble} \rho_{\rm ICM}v_{\rm ICM}^2
\label{eq:energy}
\end{equation} 
where $g$ depends on the bubble's geometry and should be of order
unity for spherical bubbles.  This energy is extracted from the wave
and available for dissipation in the ICM.

In order to verify the dimensional analysis that leads to
eq.~(\ref{eq:energy}) and to calibrate the coefficient $g$, we
undertook a set of numerical hydrodynamic simulations of shock-bubble
interactions.  We shall present the details of our numerical study
before discussing the results.

\section{Numerical Method}
\label{sec:numerical}
We performed multiple parameter studies of shock-bubble interactions.
The setup consisted of a hot bubble of varying geometry embedded in a
colder, denser medium.  Shocks or non-linear sound waves of varying
finite width $\lambda$ and pulse height (i.e., Mach number) were
introduced to travel across the bubble.

The simulations were carried out using the publically available FLASH
code \citep{fryxell:00}.  FLASH is an adaptive mesh refinement
hydrodynamics code that uses a second order accurate piece-wise
parabolic solver.  Most of our simulations were carried out on a 2D
Cartesian grid with an effective grid size of 2048x3072 cells and an
effective resolution of 64 cells across the bubble.

Figure 1 shows a typical setup with a density contrast of 100 and
pressure balance across the bubble, the ratio of sound speeds is 10
from inside to outside of the bubble.  We assumed an adiabatic
equation of state except for the shock, which is handled implicitly by
a shock capturing scheme.  The ratio of specific heats for both hot
and cold fluids was $\gamma = 5/3$.  While this will not affect the
results qualitatively, a softer $\gamma=4/3$ equation of state would
increase the compressibility of the bubble and lead to a slightly
higher value of $g$.

In most of the simulations the shock was taken to be a semi-infinite
piston with a sharp discontinuity in pressure, density and velocity,
satisfying the Rankine-Hugoniot shock jump conditions,
\citealt{landau:87}.  In the case of finite pulse width $\lambda$
(Fig.~2), the initial pulse profile was taken to be a top-hat function
in all fluid variables.

Our choice of grid implies a cylindrical bubble geometry.  To confirm
that our conclusions are independent of geometry we ran test-cases of
spherical bubbles on a 2D axi-symmetric grid.  A resolution study
confirmed that our simulations are independent of numerical
resolution.  The results are insensitive to the bubble/environment
density ratio $\rho_{\rm bubble}/\rho_{\rm ICM}$ as long as $\rho_{\rm
bubble} \ll \rho_{\rm ICM}$.  While the enforced symmetry of our 2D
simulations might slightly alter the amount of vorticity generated and
thus the numerical coefficient $g$, it has been shown that the RMI
does operate in 3D and the results of this paper will most likely not
be affected severely by restricting the analysis to 2D.

The rotational component of the velocity field was extracted by
solving the vorticity equation $\nabla \times {\mathbf v} = \nabla
\times (\nabla \times {\mathbf A})$ for the vector potential ${\mathbf
A}$ (which reduces to Poisson's equation for $-A$ in 2D Cartesian
coordinates), subject to the boundary condition of vanishing
rotational velocity at infinity, making $v_{\rm rot}$ invariant under
Galilean transformation.  We used a Fast Fourier Transform method to
solve the equation numerically.  We then calculated the kinetic energy
$E_{\rm rot} = \int dV \rho (\nabla \times {\mathbf A})^2/2$ contained
in the rotational part of the flow.  The potential velocity field can
be found by solving Poisson's equation $\nabla^2 \phi = \nabla
{\mathbf v}$.

\section{Discussion}
\label{sec:results}
\subsection{Numerical calibration of $g$}
These simulations can be used to test eq.~(\ref{eq:energy}) and to
calibrate the numerical coefficient $g$.  Fig.~2 shows the kinetic
energy $E_{\rm rot}$ in rotational flow as a function of $v_{\rm ICM}$
in the limit $R\ll\lambda$.  The numerical results are in good
agreement with the estimate from eq.~(\ref{eq:energy}), with a
numerical fit of $g=0.97$.  Fig.~2 also shows that for $\lambda
\lesssim 10R$, $g$ is proportional to $\left(\lambda/R\right)^2$.
This dependence of $g$ on $\lambda/R$ is linked to the degree of
bubble deformation produced during shock passage.  Because the total
amount of energy in the wave is proportional to $\lambda$, the energy
transfer efficiency is $\eta=g(R/\lambda)\propto\lambda/R$.  Thus,
small bubbles are more efficient at extracting energy than large ones.
\begin{figure}
\begin{center}
\resizebox{\columnwidth}{!}{\includegraphics{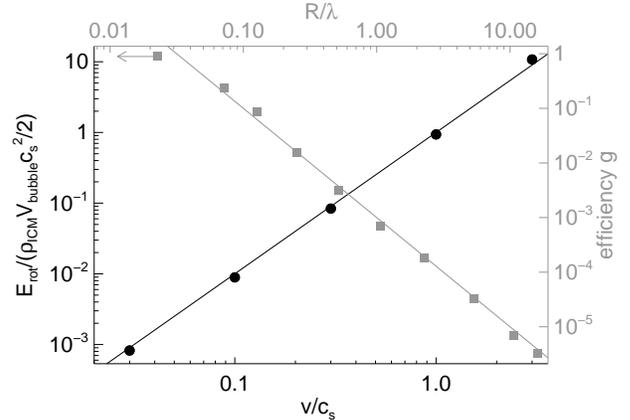}}
\end{center}
\caption{Black: Kinetic energy in the rotational velocity field
  extracted from sound/shock wave as a function of the bulk velocity
  of the shocked gas, $v_{\rm ICM}$.  Points correspond to Mach
  numbers $M=1.02,1.07,1.22,1.87, 4.20$.  Grey: Energy extraction
  efficiency $g$ as function of pulse width $\lambda$ (relative to
  maximal value $g=0.97$ for $\lambda \rightarrow \infty$).
  \label{fig:velocity}}\vspace*{-12pt}
\end{figure}

Bubble shape and orientation also affect $g$: Fig.~3 shows that
$E_{\rm rot}$ depends on $w$ and $l$, the bubble's dimensions parallel
and perpendicular to the wave.  The numerical fit for aspect ratios
$l/w$ not too far from unity is $g\propto(l/w)^{0.8}$: the {\em more}
surface area the bubble presents to the wave relative to its volume,
the {\em less} efficient the mechanism becomes.  The dependence of $g$
on filament orientation shown in Fig.~3 (grey) is due to the effective
aspect ratio changing with angle. This dependence can be easily
understood considering that all effects essential for deposition of
rotational energy into the ICM (vorticity generation and bubble
deformation) scale with the length $l$ of the filament perpendicular
to the front.
\begin{figure}
\begin{center}
\resizebox{\columnwidth}{!}{\includegraphics{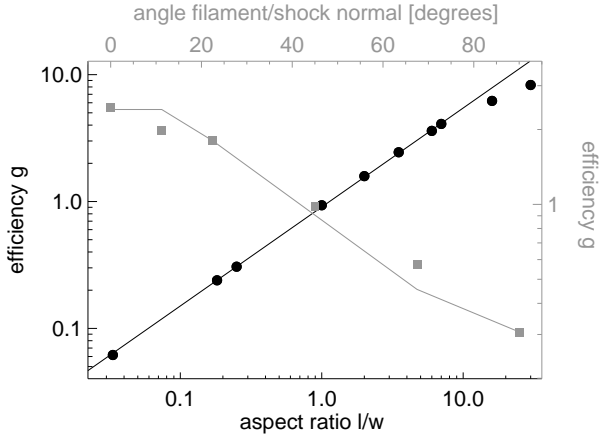}}
\end{center}
\caption{Extraction efficiency $g$ for kinetic energy in rotational
  flow as a function of aspect ratio $l/w$ (black circles) of the
  filament and best fit regression of $g\propto (l/w)^{0.8}$ (black
  line).  Bubbles exposing {\em less} surface area to the shock are
  {\em more} efficient at extracting energy per unit bubble volume.
  Also shown: Efficiency $g$ as a function of angle between filament
  and shock normal for an aspect ratio of $l/w=3.5$ (grey squares) and
  expected dependence $g \propto (\Delta y/\Delta x)^{0.8}$ from
  measured x- and y-cuts across the filament (grey line).
  \label{fig:aspectratio}}\vspace*{-12pt}
\end{figure}

\subsection{Application to galaxy clusters}
These results imply that a distribution of bubbles of hot gas whose
characteristic size is small compared to their distance $r$ from the
cluster center can extract energy from a sound or shock wave passing
over them rather efficiently.  The total energy captured equals the
energy density in the wave times the volume filled by bubbles.  To
demonstrate this, Fig.~4 shows a simulation of a random distribution
of bubbles (filling about 10\% of the volume) before and after the
shock passage.  The top panels show the creation of a turbulence field
around the bubbles in the wake of the shock.  The bottom panel shows
the decomposition of the velocity field into rotational and potential
flow.  $E_{\rm rot}$ and the viscous dissipation rate are well
localized around the bubbles even long after the shock passage.
\begin{figure}
\begin{center}
\resizebox{\columnwidth}{!}{\includegraphics{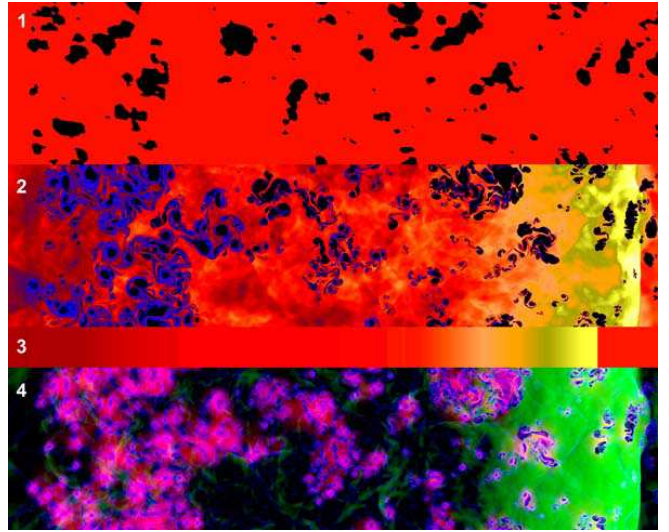}}
\end{center}
\caption{Random distribution of low density bubbles (filling fraction
  $f=10\%$) before (panel 1) and after (panel 2) passage of an
  initially square pulse of Mach 1.87, which induces a complex
  turbulence field in the wake of the bubbles and distorts the wave
  front, compared to the same setup but with uniform density (panel
  3). Panel 4: Kinetic energy of the decomposed velocity field:
  rotational flow (red); potential flow (sound/shock waves, green);
  and viscous dissipation rate (blue, in units of specific
  viscosity). Note that the rotational energy is localized around
  regions of enhanced bubble number density and that some of the
  energy in potential flow is radiated away in sound
  waves.\label{fig:random_dens}}\vspace*{-12pt}
\end{figure}

The velocity in the differentially rotating vortices drops with
distance from the bubble center and a number of dissipative processes
(such as the magneto-rotational instability \citep{balbus:91},
turbulence, and microscopic viscosity) will dissipate the energy
contained in the vortex.  The dissipated heat and entropy will go
directly into the thermal gas surrounding the bubble.  Since small
bubbles rise through the cluster gas with speeds much smaller than the
sound speed \citep{churazov:02}, the deposition of rotational energy
and the subsequent heating occur at an essentially fixed location in
the cluster compared to the passing wave.

Whether the condition $R\ll\lambda$ is satisfied in galaxy clusters
depends on the unknown size distribution of bubbles in the ICM.  In
sub-sonic models of ICM heating \citep{begelman:01,churazov:02} it is
typically assumed that a large number of small bubbles is released
rather than few large bubbles.  X-ray and radio images of nearby
clusters indicate that the distribution of radio plasma can be rather
complex \citep{owen:00,young:02b,fabian:03b}, though resolution limits
and the line of sight projection in X-ray and radio maps make a direct
detection of small filaments impossible.  The bubbles detected
directly in these clusters are of similar size to the sound waves
seen, implying a low energy extraction efficiency $g$.  In addition to
these prominent, large radio bubbles, radio observations of clusters
also show more amorphous radio halos \citep[e.g.][]{pedlar:90},
extending over the central $\sim$ 100 kpc.  Such halos {\em could}
indicate that smaller radio bubbles/filaments are present in the ICM.

Dynamical (Rayleigh-Taylor and Kelvin-Helmholtz) instabilities
\citep{landau:87} lead to fragmentation and shredding of individual
bubbles, producing a spectrum of smaller filaments.  In the absence of
magnetic fields, buoyantly rising bubbles are shredded into smaller
structures while passing a distance comparable to their own diameter,
within a time scale $\tau_{\rm shred} \sim 2 \times 10^7{\rm\,yrs}
\sqrt{(R/10\,{\rm kpc})(4\,{\rm keV}/T)}$ for a cluster of temperature
$T$.  As long as $\tau_{\rm shred}$ is comparable to or smaller than
the buoyant rise time $\tau_{\rm buoy} \sim 5\times \tau_{\rm
shred}{\left[(r_{\rm cool}/100\,{\rm kpc})(10\,{\rm
kpc}/R)\right]}^{3/2}$ a significant fraction of a bubble's volume
will be shredded within the cluster's cooling radius $r_{\rm cool}$.
A more quantitative evaluation of bubble shredding, including the
effects of the magnetic fields and viscosity, which can significantly
suppress the shredding efficiency, is a topic of ongoing research
\citep{reynolds:05,kaiser:05} and beyond the scope of this paper.

Shredding bubbles to smaller sizes increases the efficiency $g$ until
they reach a characteristic size much smaller than the wave length of
the perturbations.  After this point, $g$ is unaffected by bubble
size, though the buoyant rise velocity still decreases with bubble
size, thus leaving more and more time for dissipation processes to act
on the vorticity field.  Ultimately, if bubbles get shredded to small
enough sizes their relativistic gas content might get microscopically
mixed with the ICM.  The energy contained in the low energy electrons
can then also contribute to ICM heating via Coulomb losses.

As a possible application of the process suggested above, consider the
Perseus cluster, one of the brightest cooling flow clusters. The size
of the prominent central radio bubbles is about 10 kpc. This sets the
typical thickness of $\lambda\sim$10 kpc of the waves generated by
bubble inflation, consistent with the concentric ripples seen in X-ray
images \citep{fabian:03b}.  Since the bubbles are comparable in size
to $\lambda$, they are inefficient at extracting energy from AGN
induced sound waves.  However, if the ICM in Perseus is filled with
numerous smaller bubbles, the typical attenuation length for a
compression wave will be $L\sim\lambda/f$, where $f$ is the fraction
of cluster volume filled by bubbles.  For $f \sim 10$\%, most of the
wave energy would be trapped in the inner $\sim$100 kpc of the
cluster, where most of the excess heating is required.  Values of $f$
up to 10\% might be reasonable if a significant fraction of the radio
plasma seen in cluster cavities with $R\sim \lambda$
\citep{birzan:04,giovannini:04} is shredded to smaller scales, or if
the radio plasma producing the diffuse cluster radio halo is
filamentary rather than homogeneous.

Thus, even in the absence of microscopic viscosity, a bubbly ICM can
prevent leakage of wave energy out of the central region.  Moreover,
bubbles will also boost the dissipation of sound waves created by
other mechanisms (e.g., cluster collisions) in the central region,
where $f$ is presumably largest.  Generally, filling factors of $f\sim
few\% - few \times 10\%$ are necessary to provide the required heating
in cooling flow clusters.

\subsection{Strong shocks in the cluster center}
Recent deep cluster observations have revealed large scale surface
brightness discontinuities that are consistent with moderate strength
shocks \citep{mcnamara:05,nulsen:05}.  This could mean that cluster
centers are subject to much stronger shocks than the cool X-ray rims
around radio bubbles would imply.  The question of why strong, jet
driven shocks have not been found has been a puzzle ever since {\em
Chandra} began observing the temperature structures of the inner
regions of galaxy clusters.  The presence of bubbles in the ICM might
help explain the lack of observational evidence for such shocks: the
large sound speed inside the bubbles means that they rapidly broadcast
the arrival of a shock, which locally jumps ahead when passing over a
filament.  This not only scatters some of the shock into isotropic
sound waves, it also distorts and broadens the front (Fig.~4). When
observing such a shock in X-rays it will appear weaker than a coherent
shock in the absence of bubbles.

Such strong shocks might be efficient enough to heat the cluster
directly.  However, the presence of bubbles increases the dissipation
efficiency {\em regardless} of shock strength.  Therefore, the
proposed mechanism would still operate.

\section{Conclusions}
\label{sec:conclusions}
The presence of bubbles of relativistic gas in galaxy clusters can
provide a significant boost to the efficiency with which shock and
sound waves energize the gas.  Given the ubiquity of both sound waves
and of relativistic gas in galaxy clusters, this may well be a
significant (if not dominant) channel of cluster gas heating.  More
generally, our results demonstrate the high efficiency of the RMI,
which is known to have a broad range of scientific and industrial
applicability.
\vspace*{6pt}

{\noindent \bf Acknowledgements:} We would like to thank Emily
Levesque and Thomas Janka for helpful discussions. SH acknowledges
support by the National Aeronautics and Space Administration through
Chandra Postdoctoral Fellowship Award Number PF3-40026 issued by the
Chandra X-ray Observatory Center, which is operated by the Smithsonian
Astrophysical Observatory for and on behalf of the National
Aeronautics Space Administration under contract NAS8-39073.

\end{document}